\documentclass[aps,prb,twocolumn,showpacs,superscriptaddress]{revtex4}

\usepackage{graphicx}
\usepackage{amsmath}

\begin{document}

\title{Gap and magnetic engineering via doping and pressure in tuning the colossal magnetoresistance in (Mn$_{1-x}$Mg$_x$)$_3$Si$_2$Te$_6$}
\author{Chaoxin Huang}
\author{Mengwu Huo}
\author{Xing Huang}
\author{Hui Liu}
\author{Lisi Li}
\affiliation{Center for Neutron Science and Technology, Guangdong Provincial Key Laboratory of Magnetoelectric Physics and Devices, School of Physics, Sun Yat-Sen University, Guangzhou 510275, China }
\author{Ziyou Zhang}
\author{Zhiqiang Chen}
\affiliation{Center for High Pressure Science and Technology Advanced Research, Shanghai 201203, China }
\author{Yifeng Han}
\affiliation{Center for Materials of the University, School of Molecular Sciences, Arizona State University, Tempe, Arizona 85287, USA }
\author{Lan Chen}
\author{Feixiang Liang}
\affiliation{Center for Neutron Science and Technology, Guangdong Provincial Key Laboratory of Magnetoelectric Physics and Devices, School of Physics, Sun Yat-Sen University, Guangzhou 510275, China }
\author{Hongliang Dong}
\affiliation{Center for High Pressure Science and Technology Advanced Research, Shanghai 201203, China }
\author{Bing Shen}
\affiliation{Center for Neutron Science and Technology, Guangdong Provincial Key Laboratory of Magnetoelectric Physics and Devices, School of Physics, Sun Yat-Sen University, Guangzhou 510275, China }
\author{Hualei Sun}
\email{sunhlei@mail.sysu.edu.cn}
\affiliation{School of Science, Sun Yat-Sen University, Shenzhen 518107, China }
\author{Meng Wang}
\email{wangmeng5@mail.sysu.edu.cn}
\affiliation{Center for Neutron Science and Technology, Guangdong Provincial Key Laboratory of Magnetoelectric Physics and Devices, School of Physics, Sun Yat-Sen University, Guangzhou 510275, China }

\begin{abstract}

Ferrimagnetic nodal-line semiconductor Mn$_3$Si$_2$Te$_6$ keeps the records of colossal magnetoresistance (CMR) and angular magnetoresistance (AMR). Here we report tuning the electronic transport properties via doping and pressure in (Mn$_{1-x}$Mg$_x$)$_3$Si$_2$Te$_6$. As the substitution of nonmagnetic Mg$^{2+}$ for magnetic Mn$^{2+}$, ferrimagnetic transition temperature $T_C$ gradually decreases, while the resistivity increases significantly. At the same time, the CMR and AMR are both enhanced for the low-doping compositions (e.g., $x = 0.1$ and 0.2), which can be attributed to doping-induced broadening of the band gap and a larger variation range of the resistivity when undergoing a metal-insulator transition by applying a magnetic field along the $c$ axis. On the contrary, $T_C$ rises with increasing pressure due to the enhancement of the magnetic exchange interactions until a structural transition occurs at $\sim$13 GPa. Meanwhile, the activation gap is lowered under pressure and the magnetoresistance is decreased dramatically above 6 GPa where the gap is closed. At 20 and 26 GPa, evidences for a superconducting transition at $\sim$5 K are observed. The results reveal that doping and pressure are effective methods to tune the activation gap, and correspondingly, the CMR and AMR in nodal-line semiconductors, providing an approach to investigate the magnetoresistance materials for novel spintronic devices.

\end{abstract}
\maketitle

\section{INTRODUCTION}	

Since giant magnetoresistance was discovered in magnetic multilayers, the spin-charge coupling has been widely investigated for decades in a number of fields, including magnetic storage, magnetic sensor, magnetometer and so on\cite{grunberg1986,baibich1988,xiao1992,thompson2008,ennen2016,dieny1994}. CMR is more sensitive to a magnetic field than giant magnetoresistance, whose variation in resistance can be close to 100\%. CMR was first reported in doped manganese perovskite, where the double exchange mechanism and dynamic Jahn-Teller effect were demonstrated to play a key role\cite{von1993,jin1994,millis1996}. Other mechanisms were suggested for the later discovered CMR materials, such as strong magnetic fluctuations in EuCd$_2$P$_2$\cite{wang2021}, magnetic polarons in Ti$_2$Mn$_2$O$_7$\cite{shimakawa1996,majumdar1998} and Eu$_5$In$_2$Sb$_6$\cite{rosa2020}, and spin-orbit coupling/spin-texture driven CMR in EuTe$_2$ and EuMnSb$_2$\cite{yin2020,yang2021,sun2021}. Two-dimensional (2D) layered materials have higher spatial anisotropy, which is a fertile ground to explore new material systems with CMR and AMR\cite{ogasawara2021,sun2021}.

Indeed, the quasi-2D ferrimagnetic (FIM) semiconductor Mn$_3$Si$_2$Te$_6$ was found to exhibit the largest CMR and AMR, where the resistivity decreases by 10$^7$ as defined by ($\rho$$_H$-$\rho$$_0$)/$\rho$$_H$ for $H\|c$ axis, but changes slightly in resistivity when $H$ is along the $ab$ plane\cite{vincent1986,ni2021,seo2021,sala2022,kwon2023,mijin2023,lovesey2023}. It was suggested that the nodal-line degeneracy of Mn$_3$Si$_2$Te$_6$ can be controlled by spin orientation. When the applied field is along the $c$ axis, spin-orbit coupling (SOC) between Mn and Te will result in band splitting and one of the bands will shift toward the Fermi level, leading to the closing of the electric gap and metal-insulator transition (MIT)\cite{seo2021}. Later, chiral orbital currents (COC) were suggested to circulate along the edges of MnTe$_6$ octahedra. In this scenario, the coupling between the spin of Mn$^{2+}$ and the effective moments of the COC could reduce the electron scattering when a magnetic field is applied along the $c$ axis\cite{zhang2022Control}. However, single-crystal neutron diffraction and magnetization measurements reveal that the applied field along the $c$ axis smoothly tilts the magnetic moment during the CMR occurs \cite{ni2021,seo2021,ye2022}. The activation gap may be a key gradient in the mechanism. Thus, further research to improve the magnitudes of the CMR and AMR and elucidate the mechanism is crucial to promote potential applications in the future.

Doping and pressure are considered as effective methods to tune quantum transport properties in electronic correlated materials\cite{zhang2022,zhang2023,sun2023coexistence,sun2023exchange,sun2021magnetism,Cai2020}. In this work, we report the successful synthesis of a series of (Mn$_{1-x}$Mg$_x$)$_3$Si$_2$Te$_6$ single crystals from $x$ = 0 to 1.0 at 0.1 intervals and investigations of the evolutions of structure, magnetism, and electric transport properties with doping and pressure. We find the Mg-doping can suppress the FIM, increase the resistivity by enlarging the activation gap, and enhance the CMR and AMR by about two orders of magnitude compared to the undoped compound. 
On the other hand, pressure induces the closure of the thermal activation gap at $\sim$7.0 GPa accompanied by the disappearance of the CMR, a structural transition at 14.6 GPa, and a superconducting transition at 5 K and 20 GPa.
Our results reveal that the nodal-line topological band gap can be widened or narrowed by doping or applying pressure, thus tuning the CMR and AMR to a required magnitude.

\section{EXPERIMENTAL DETAILS}

Single crystals of (Mn$_{1-x}$Mg$_x$)$_3$Si$_2$Te$_6$ ($0\le x\le1$) were grown by the self-flux method and characterized by x-ray diffraction (XRD, Empyrean), energy dispersive x-ray spectroscopy (EDS, EVO Zeiss), Laue diffractometer(Photonic Science), and physical property measurement system (PPMS, Quantum Design). For partial samples with resistance exceeding the measuring range (e.g., data for $x$ = 0.1 at low temperature and low field), an external voltammeter (Keithley) was employed while necessary. 

High-pressure electrical transport measurements of Mn$_3$Si$_2$Te$_6$ single crystals were carried out using a miniature diamond anvil cell (DAC) made from a Be-Cu alloy on the PPMS. Diamond anvils with a 300 $\mu$m culet were used. The corresponding sample chamber with a 110 $\mu$m diameter was made in an insulating gasket achieved by cubic boron nitride and epoxy mixture. KBr powders were employed as the pressure-transmitting medium, providing a quasi-hydrostatic environment. The pressure was calibrated by measuring the shift of the fluorescence wavelength of the ruby sphere, which was loaded in the sample chamber. The standard four-probe technique was adopted for these measurements. 

The in situ high-pressure synchrotron powder XRD patterns were collected at 300 K with an x-ray wavelength of 0.6199 {\AA} on the Shanghai Synchrotron Radiation Facility. A symmetric DAC with a pair of 300 $\mu$m diameter culet was used. The sample chamber was drilled by laser with a diameter of 120 $\mu$m. Daphne oil was used as the pressure-transmitting medium, and the pressure in the DAC was also calibrated by the shift of the fluorescence of the ruby sphere.
The high-pressure XRD data were initially integrated using Dioptas\cite{Clemens2015} (with a CeO2 calibration) and fitted using the Pawley method in Topas-Academic V6 software\cite{Pawley1981}.

\section{RESULTS AND DISCUSSION }	

XRD measurements reveal that the crystal structure of (Mn$_{1-x}$Mg$_x$)$_3$Si$_2$Te$_6$ ($0\le x\le1$) preserves the trigonal space group $P\overline{3}1c$. A small number of impurities of MnTe$_2$ can be identified in the XRD patterns of $x$ = 0 and 0.4 compounds\cite {xu2018}. The lattice parameters are smoothly enlarged as the Mg doping. The nominal compositions are close to the EDS-determined compounds [Supplementary Fig. S1]. We use the nominal $x$ to indicate the composition of (Mn$_{1-x}$Mg$_x$)$_3$Si$_2$Te$_6$ through out this paper.

\begin{figure}[t]
\includegraphics[scale=0.38]{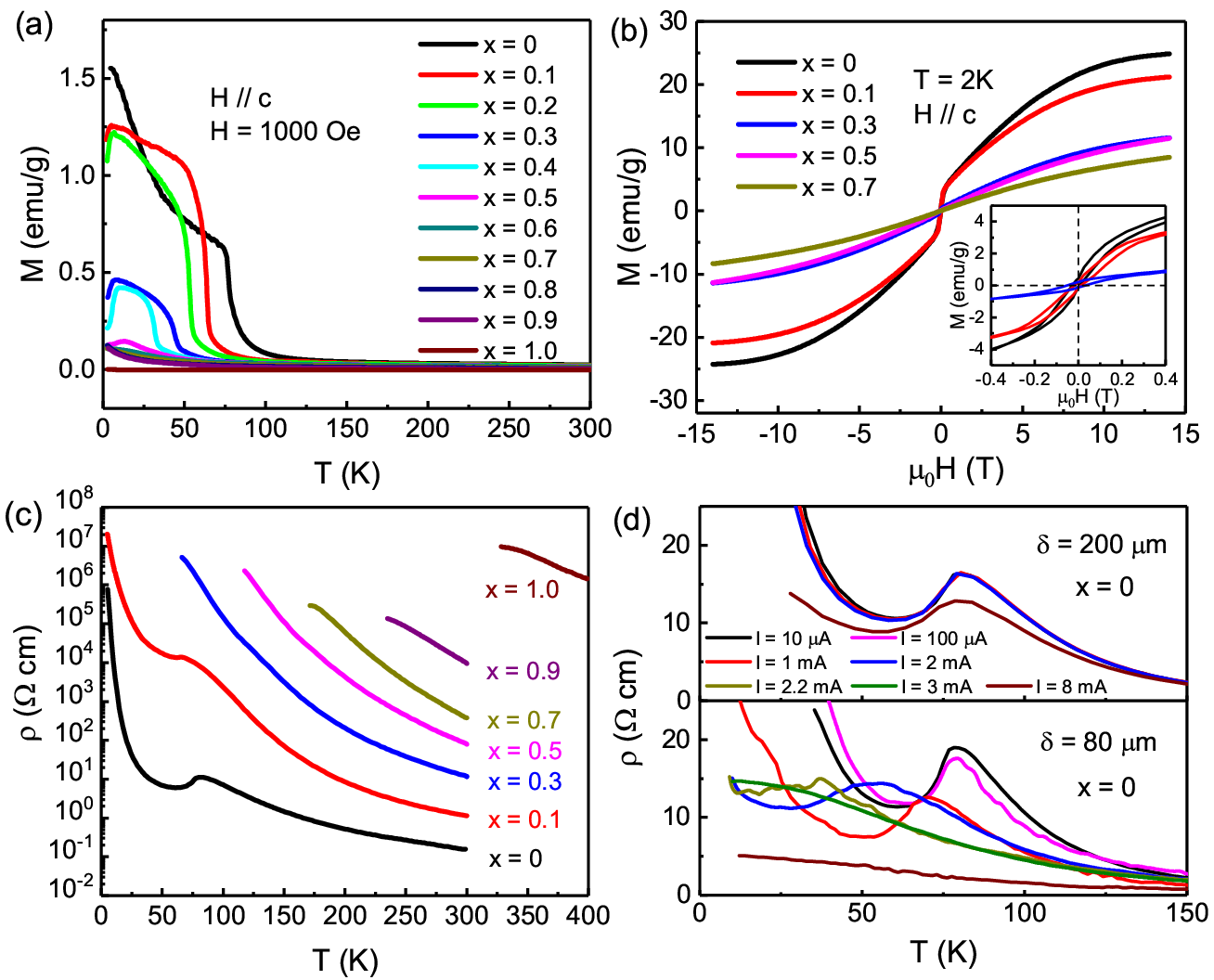}
\caption{(a) Temperature dependence of magnetization of (Mn$_{1-x}$Mg$_x$)$_3$Si$_2$Te$_6$ with the magnetic field parallel to the $c$ axis and $H$ = 1000 Oe. (b) Temperature dependence of magnetization of (Mn$_{1-x}$Mg$_x$)$_3$Si$_2$Te$_6$ at 2 K with an applied field parallel to the $c$ axis from -14 to 14 T. The inset shows the hysteresis loops of the $x$ = 0, 0.1, and 0.3 samples at low fields. (c) Temperature dependence of resistivity of (Mn$_{1-x}$Mg$_x$)$_3$Si$_2$Te$_6$. The resistivity of the $x$ = 1.0 sample was measured from 328 to 400 K due to the large magnitude which exceeded the measurement range at lower temperatures. (d) Temperature dependence of resistivity of Mn$_3$Si$_2$Te$_6$ with thicknesses of 200 and 80 $\mu$m measured with various currents.}
\label{fig1}
\end{figure}

Figure \ref{fig1} shows the doping effects on magnetization and resistivity of (Mn$_{1-x}$Mg$_x$)$_3$Si$_2$Te$_6$. The temperature dependence of magnetization in Fig. \ref{fig1}(a) reveals a transition at $\sim$78 K for Mn$_3$Si$_2$Te$_6$ that is consistent with the reported FIM order\cite{may2017,liu2018,martinez2020,liu2021polaronic}. The FIM order is progressively suppressed with increasing doping until the disappearance for $x>0.5$. In addition to the FIM transition, an abnormal decline on the magnetization curve can be observed at low temperatures for $0.1\le x\le0.5$, which could be ascribed to a spin reorientation transition due to a doping-induced magnetic dilution (Supplementary Fig. S2)\cite{Liu2023}. The saturation magnetization decreases with increasing $x$ [Fig. \ref{fig1}(b)]. The hysteresis loops resulting from the FIM can be observed in the low field range in both the undoped and doped samples [see inset in Fig. \ref{fig1}(b)]. 

 \begin{figure}[t]
\includegraphics[scale=0.51]{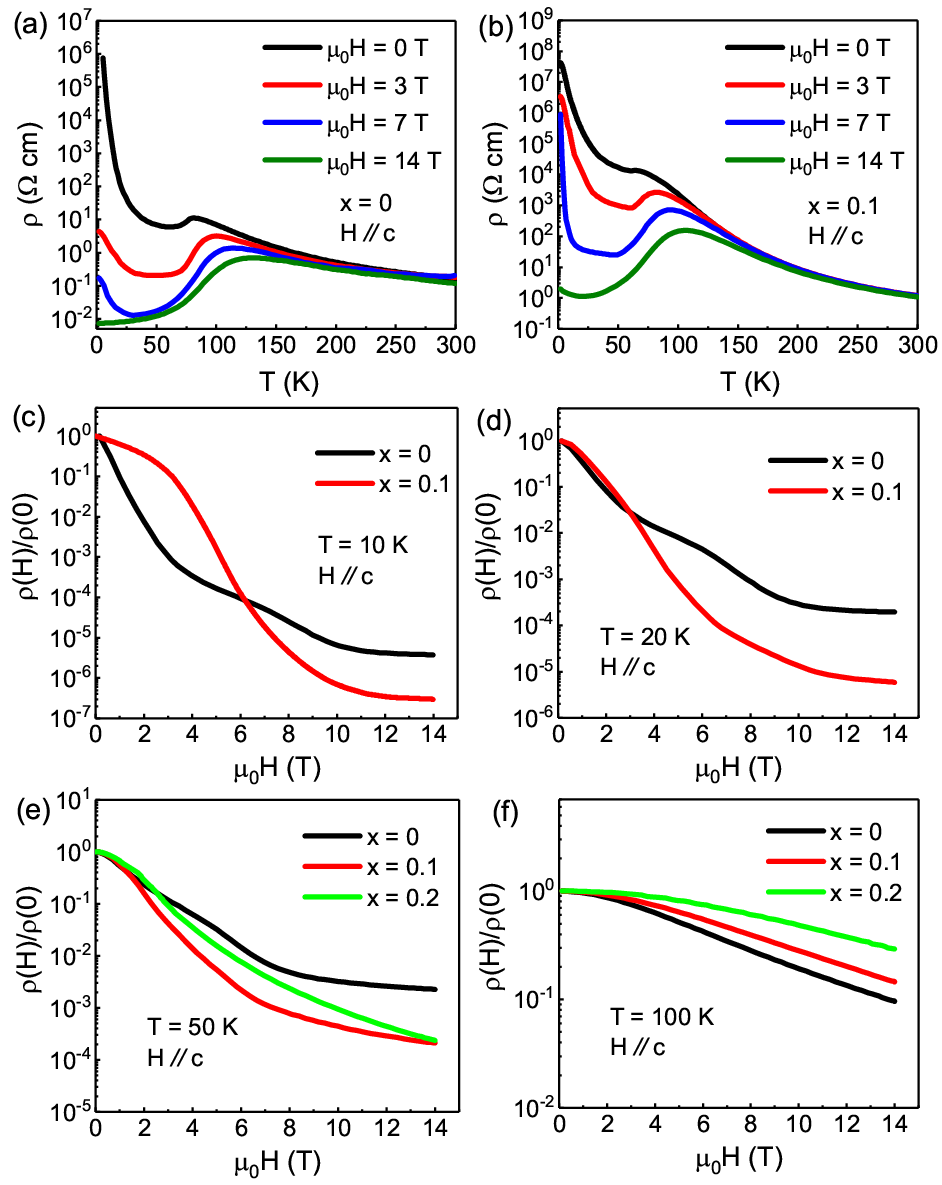}
\caption{(a) Resistivity as a function of temperature in different magnetic fields parallel to the $c$ axis for the undoped sample and (b) for the $x$ = 0.1 sample. (c) Field dependence of $\rho$$_H$/$\rho$$_0$ for the samples of $x$ = 0 and 0.1 at 10 K, (d) 20 K, and for the samples of $x$ = 0, 0.1, and 0.2 at (e) 50 K and (f) 100 K.}
\label{fig2}
\end{figure}

Temperature dependence of the resistivity for different compositions is displayed in Fig. \ref{fig1}(c). Anomalies on the resistivity of $x$ = 0 and 0.1 samples correspond to the $T_C$s. When the $x$ exceeds 0.1, the anomaly associated with $T_C$ cannot be observed due to the limit of the measured resistance range. The values of the resistivity increase from 10$^{-1}$ to 10$^{7}$ $\Omega$ cm at $\sim$300 K in the doping range of $0\le x\le1$. The resistivity for Mg$_3$Si$_2$Te$_6$ is too large to be measured below 328 K in our experimental setup\cite{huang2023}. We fit the resistivity above $T_C$ using the polaron hopping model\cite{liu2021polaronic} $\rho$($T$) = $AT$exp($E$$_a$/$k$$_B$$T$), where $E_a$ is the activation energy and $k_B$ is the Bolzman constant. The results reveal linearly increased $E_a$ from 83.5 to 408.6 meV with the Mg$^{2+}$ doping [Fig. \ref{fig5}], revealing that the gap changes uniformly as doping, which may facilitate the electronic structure tuning in real applications. Figure \ref{fig1}(d) shows the resistivity measurements on Mn$_3$Si$_2$Te$_6$ single crystals with different currents and thicknesses. The $T_C$ can be tuned by the magnitude of current for the single crystal with a thickness of $\sigma$ = 80 $\mu$m, consistent with the previously proposed COC state\cite{zhang2022Control}. However, the anomalies in resistivity for the $\sigma$ = 200 $\mu$m sample are hardly changed until the current $I$ is up to 8 mA. According to the discrepancy of the $T_C$s for different currents and samples, the COC state does not depend on the current magnitude but the current density. The critical current density for the FIM state at 0 K is estimated to be 95 mA/mm$^2$ [Supplementary Fig. S1(d)].

To reveal the doping effect on the CMR, we show the temperature dependence of resistivity with magnetic fields parallel to the $c$ axis for the undoped $x=0$ and doped $x=0.1$ compounds in Figs. \ref{fig2}(a) and \ref{fig2}(b). They all exhibit large CMR below the $T_C$. Similar measurements on the $x=$ 0.2, 0.3, 0.5, and 0.7 compositions are shown in the Supplementary Fig. S3.
At 5 K, the resistivity of Mn$_3$Si$_2$Te$_6$ drastically decreases by about 8 orders of magnitude from 10$^6$ $\Omega$cm at zero filed to 10$^{-2}$ $\Omega$cm at $\mu_0H$ = 14 T. The change is smaller for (Mn$_{0.9}$Mg$_{0.1}$)$_3$Si$_2$Te$_6$ at 5 K. However, over a wider temperature range from 10 K to the $T_C$, the drop of resistivity for the doped $x=0.1$ compound under high magnetic fields is larger.
To compare the CMR as doping quantitatively, we plot $\rho(H)/\rho(0)$ as a function of the magnetic field at 10, 20, and 50 K below the $T_C$, and 100 K above the $T_C$ in Figs. \ref{fig2}(c-f).
Due to the limitation of the measured resistivity range, the data of the $x=0.2$ compound at 10 and 20 K are not shown in Figs. \ref{fig2}(c) and \ref{fig2}(d). The change of the magnitude of $\rho(H)/\rho(0)$ for the doped $x=0.1$ sample is about 2 orders larger than that of the undoped $x=0$ sample below $T_C$ at 14 T. The magnitudes of $\rho(H)/\rho(0)$ reverse below 6.0 T at 10 K, 2.8 T at 20 K, and 1.8 T at 50 K, respectively. The CMR effect becomes much weaker above $T_C$ as shown in Fig. \ref{fig2}(f). Thermal fluctuations, SOC, and thermal activation gap are three ingredients that govern the CMR in (Mn$_{1-x}$Mg$_x$)$_3$Si$_2$Te$_6$. Increasing the temperature below $T_C$ and strengthening the SOC will promote the conductance of electrons as expected. Our results demonstrate that substituting Mn$^{2+}$ by nonmagnetic Mg$^{2+}$ will enlarge the thermal activation gap, enhancing the CMR in this system.

\begin{figure}[t]
\includegraphics[scale=0.25]{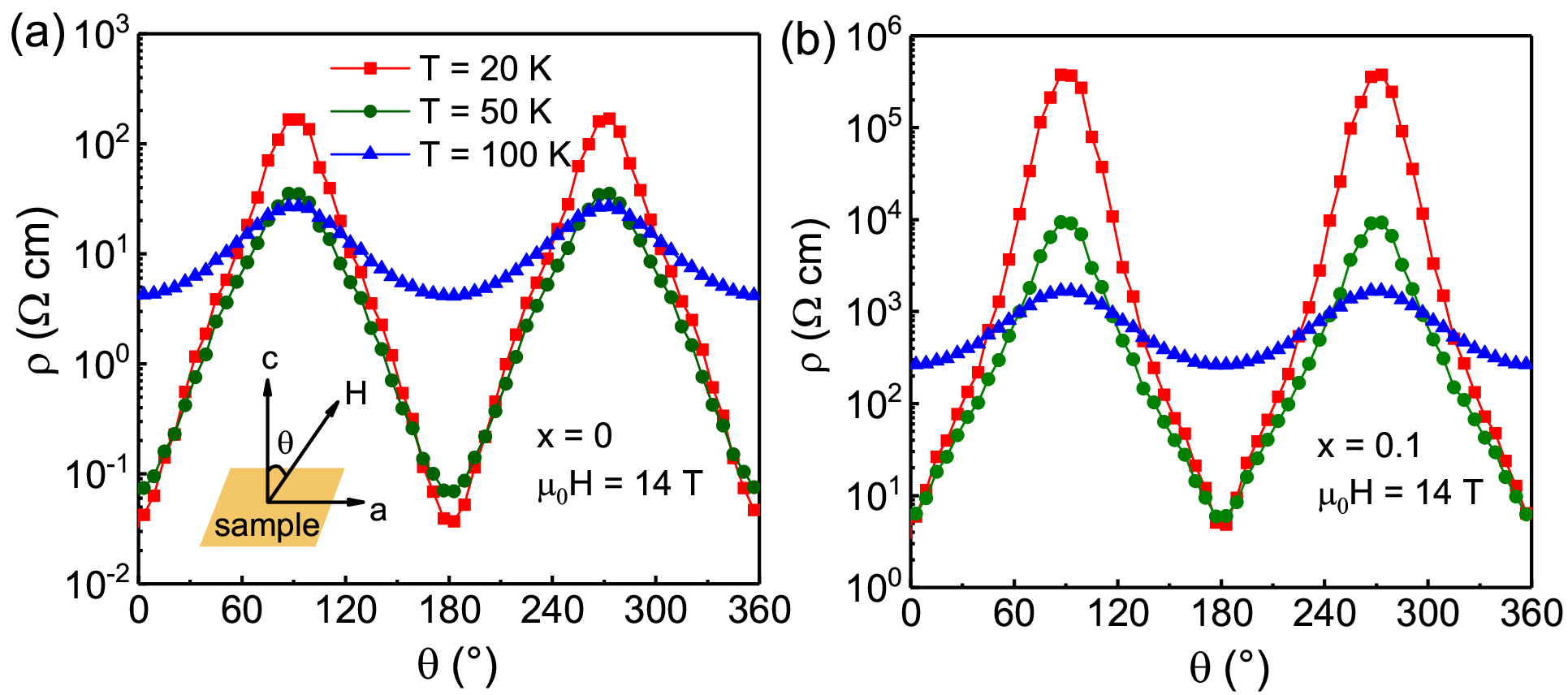}
\caption{(a) Angular dependence of resistivity for Mn$_3$Si$_2$Te$_6$ and (b) (Mn$_{0.9}$Mg$_{0.1}$)$_3$Si$_2$Te$_6$ with $\mu_0H$ = 14 T at 20, 50, and 100 K. The inset in (a) shows the direction of the magnetic field, where $\theta$ is the angle between $H$ and the $c$ axis.}
\label{fig3}
\end{figure}

Figure \ref{fig3} shows a comparison of the colossal AMR for the undoped ($x=0$) and doped ($x=0.1$) compounds at 20 and 50 K below $T_C$, and 100 K above $T_C$. The AMR exhibits a two-fold rotational symmetry. The resistivity as a function of the angle of the magnetic field to the $c$ axis evolves close to linearly, consistent with the scenario that the spins of Mn$^{2+}$ couple with the moments of the COC along the edges of the MnTe$_6$ octahedra\cite{zhang2022Control}. Upon 10$\%$ Mg doping,
the thermal activation gap is enlarged, and the magnitude of the AMR in (Mn$_{0.9}$Mg$_{0.1}$)$_3$Si$_2$Te$_6$ is enhanced an order at the same measurement conditions. Nonetheless, it is expected that the AMR will be enhanced in a wider doping range and at lower temperatures\cite{seo2021}.

\begin{figure}[t]
\includegraphics[scale=0.29]{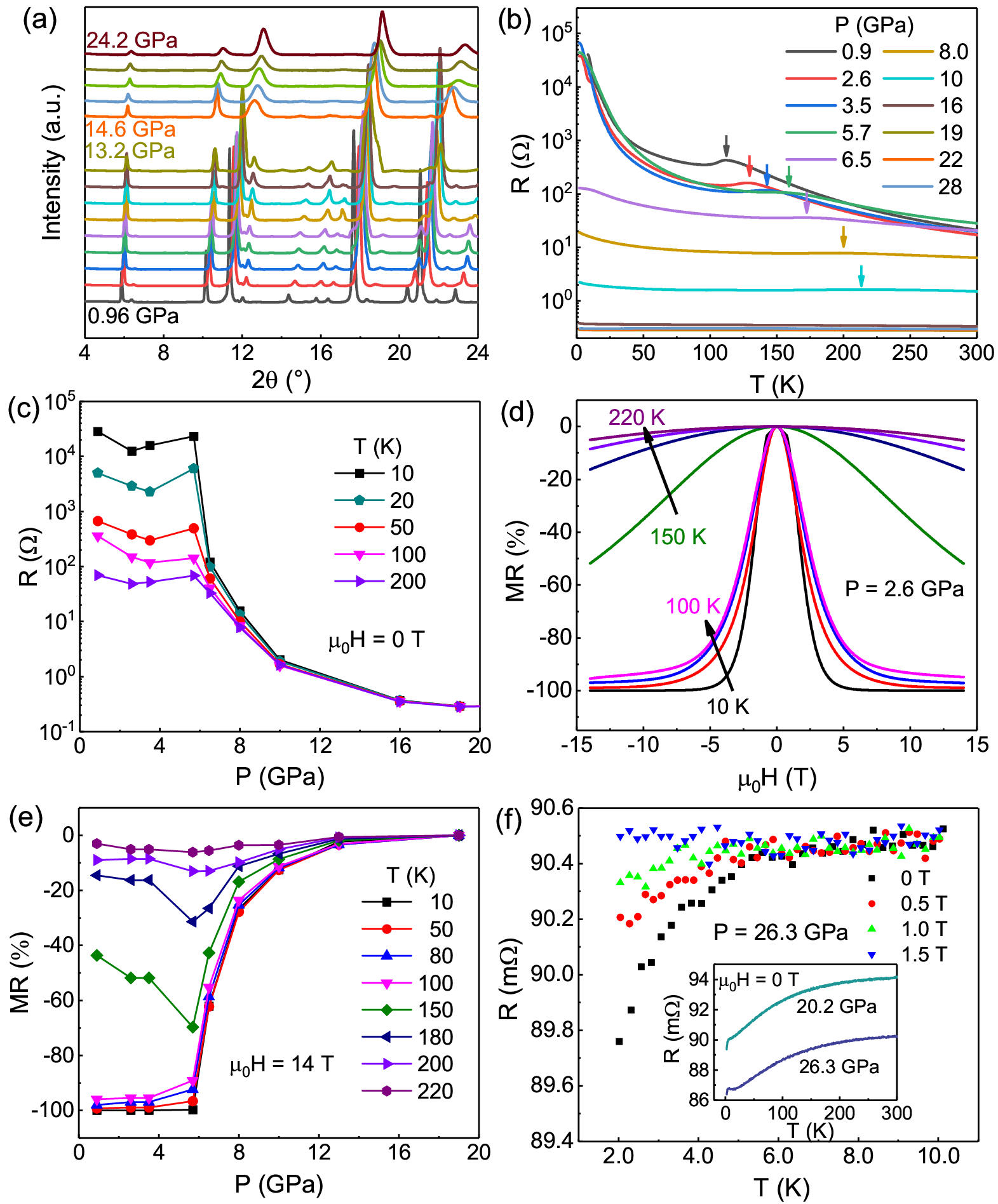}
\caption{
(a) XRD patterns of Mn$_3$Si$_2$Te$_6$ under pressures from 0.96 to 24.2 GPa. A structural transition occurs between 13.2 and 14.6 GPa. (b) Temperature dependence of resistance of Mn$_3$Si$_2$Te$_6$ at various pressures up to 28 GPa. The Curie transition temperatures $T_C$s under pressure are determined from the humps on resistance marked with the arrows.
(c) Pressure dependences of the resistance with $\mu_0H=0$ T and (e) the $MR$ with $\mu_0H=14$ T at various temperatures.
(d) $MR$ defined by ($\rho$$_H$-$\rho$$_0$)/$\rho$$_0\times$100\% at various temperatures at 2.6 GPa.
(f) Resistance measurements below 10 K at 26.3 GPa and various magnetic fields. The inset shows the date from 2 to 300 K  at 20.2 and 26.3 GPa with zero magnetic field. The drops of resistance at $\sim$5 K and the behavior under magnetic fields indicate pressure-induced superconductivity.
}
\label{fig4}
\end{figure}

\begin{figure}[t]
\includegraphics[scale=0.28]{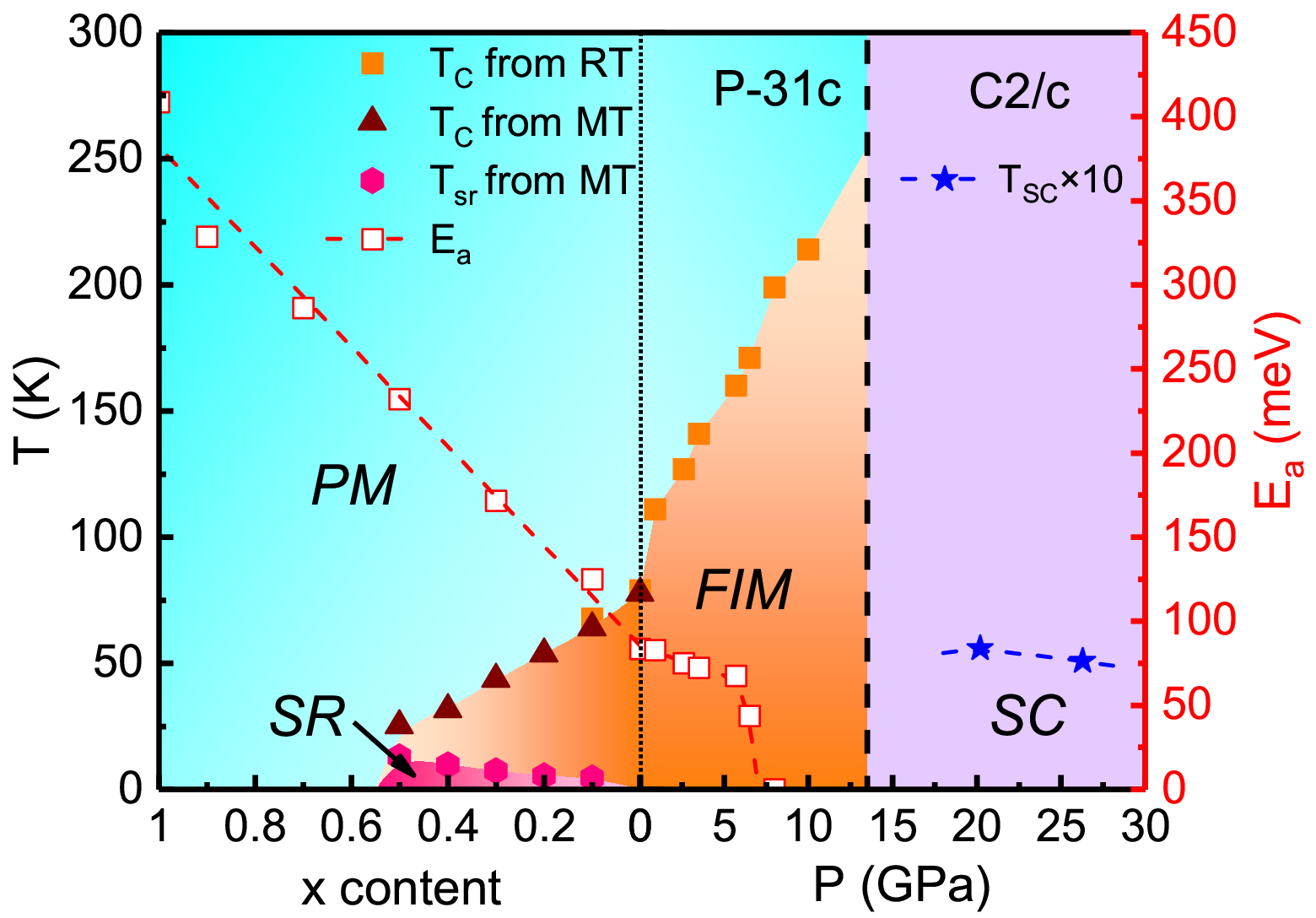}
\caption{A phase diagram of (Mn$_{1-x}$Mg$_x$)$_3$Si$_2$Te$_6$ aganist doping and pressure. The paramagnetic (PM), ferrimagnetic (FIM), spin reorientation (SR), and superconducting (SC) regions are determined by the resistance and magnetic measurements. The SC transition temperatures $T_{SC}$s have been magnified ten times for visualization. The thermal activation gaps, E$_a$s, are fitted from the resistance.
}
\label{fig5}
\end{figure}

Pressure can effectively tune the structure, electronic transport, and magnetism of layered structural materials\cite{sun2023coexistence,sun2021magnetism,olmos2023}. Figure \ref{fig4}(a) displays the high-pressure XRD patterns of Mn$_3$Si$_2$Te$_6$ up to 24.2 GPa. A clear structural transition is identified between 13.2 and 14.6 GPa, consistent with the reported Raman spectroscopy measurements\cite{wang2022}. The XRD patterns above 14.6 GPa are consistent with the space group $C2/c$ with a monoclinic structure where the lattice constants $a=5.682, b=4.080$, $c=11.559$ \AA\, and $\beta=97.88^\circ$ at 14.6 GPa. The lattice parameters as a function of pressure can be deduced from the XRD patterns (Supplementary Fig. S4). However, the accurate atomic positions cannot be determined due to the limited reflection peaks.

Under pressure, the resistance measurements reveal a semiconducting to metallic transition [Fig. \ref{fig4}(b)]. Below the pressure of the structural transition, the $T_C$ can be identified as a hump in resistance, yielding a linearly increased $T_C$ from $\sim$78 K at ambient pressure to $\sim$214 K at 10 GPa. The anomalous Hall effect of Mn$_3$Si$_2$Te$_6$ under pressure is intimately coupled with the magnetism (Supplementary Fig. S5). The thermal activation gap decreases with increasing pressure until closing at around 7 GPa. The activation gaps fitted by the polaron hopping model and the $T_C$s of Mn$_3$Si$_2$Te$_6$ under pressure have been presented in Fig. \ref{fig5}. The resistance measured at various pressures and temperatures also shows an abrupt drop at $\sim$6 GPa [Fig. \ref{fig4}(c)]. Figure \ref{fig4}(d) shows typical magnetoresistance ($MR$ defined as ($\rho$$_H$-$\rho$$_0$)/$\rho$$_0\times$100\%) measurements at 2.6 GPa and different temperatures (for more pressures, see supplementary Fig. S6). The $MR$ shows an abrupt change across $T_C=127$ K at 2.6 GPa [Fig. \ref{fig4} (d)] and a sharp drop at $\sim$6 GPa [Fig. \ref{fig4} (e)], revealing that the SOC and thermal activation gap are the dominating ingredients that govern the $MR$ as elevated temperature and pressure, respectively. 

The resistance of Mn$_3$Si$_2$Te$_6$ at 20.2 and 26.3 GPa are shown in Fig. \ref{fig4} (f) to investigate the electronic property of the monoclinic structure. Clear drops can be observed below 5 K on resistance. By applying a magnetic field up to 1.5 T, the resistance at 26.3 GPa below 5 K is enhanced obviously and the drop is finally suppressed, suggesting a superconducting transition. We note that it is common for Te-contained compounds showing superconductivity under pressure, such as WTe$_2$\cite{pan2015,kang2015}, CrSiTe$_3$\cite{Cai2020}, and EuTe$_2$\cite{yang2022,sun2023exchange}, where the electrons of Te play a dominating role in superconductivity.

A comprehensive phase diagram combining the electronic, magnetic, and structural measurements as functions of doping and pressure is shown in Fig. \ref{fig5}. The $T_C$ of the FIM transition decreases as doping but increases as applying pressure. The evolution of the thermal activation gap, $E_a$, reverses, changing from 408.6 meV in Mg$_3$Si$_2$Te$_6$ to 83.5 meV in Mn$_3$Si$_2$Te$_6$ at ambient pressure, and closing at 7 GPa. The magnetoresistance decreases dramatically as the $E_a$ closes, revealing that the CMR of Mn$_3$Si$_2$Te$_6$ is directly coupled with $E_a$. 
The presence of magnetoresistance requires SOC that is comparable with the electronic gap. In (Mn$_{1-x}$Mg$_x$)$_3$Si$_2$Te$_6$, the spins of Mn$^{2+}$ possibly couple with the moments of the COC of the MnTe$_6$ octahedra. The superconductivity observed in the monoclinic structure under pressure may span a large pressure range.

\section{SUMMARY}

In conclusion, we systematically studied the CMR, AMR, FIM, and structures of (Mn$_{1-x}$Mg$_x$)$_3$Si$_2$Te$_6$ as functions of doping and pressure. As Mn$^{2+}$ is replaced by Mg$^{2+}$ progressively, the FIM order is suppressed gradually until disappearing for $x>0.5$. For the compositions with $x$ between 0.1 and 0.5, a spin reorientation transition appears evidenced by a magnetic anomaly on magnetization at low temperatures. The thermal activation gap is enlarged linearly by the Mg doping. The CMR and AMR are both enhanced in the low-doped samples compared to the parent, which can be attributed to a doping-induced broadening of the band gap and a larger variation range of resistivity. On the contrary, $T_C$ is enhanced with increasing pressure due to the enhanced magnetic exchange interactions until disappearing when undergoing a structural transition above 13.2 GPa. The gap closes and a MIT occurs at $\sim$7 GPa, where the CMR is abruptly reduced. A superconducting transition at 5 K in the monoclinic phase under pressure is also observed. Our finding reveals that both doping and pressure are effective methods to change the band structure and magnetism, consequently enhancing the CMR and AMR  in the FIM (Mn$_{1-x}$Mg$_x$)$_3$Si$_2$Te$_6$.

\section{ACKNOWLEDGMENTS}

Work at Sun Yat-Sen University was supported by the National Key Research and Development Program of China (grant Nos. 2023YFA1406500, 2023YFA1406002), the National Natural Science Foundation of China (grant No. 12174454), the Guangdong Basic and Applied Basic Research Foundation (grant No. 2021B1515120015), the Guangzhou Basic and Applied Basic Research Funds (grant Nos. 2024A04J6417), and the Guangdong Provincial Key Laboratory of Magnetoelectric Physics and Devices (grant No. 2022B1212010008). We also thank the BL15U1 station in Shanghai Synchrotron Radiation Facility (SSRF) for the help in characterizations. 
\bibliography{ref}

\begin{thebibliography}{46}
\expandafter\ifx\csname natexlab\endcsname\relax\def\natexlab#1{#1}\fi
\expandafter\ifx\csname bibnamefont\endcsname\relax
  \def\bibnamefont#1{#1}\fi
\expandafter\ifx\csname bibfnamefont\endcsname\relax
  \def\bibfnamefont#1{#1}\fi
\expandafter\ifx\csname citenamefont\endcsname\relax
  \def\citenamefont#1{#1}\fi
\expandafter\ifx\csname url\endcsname\relax
  \def\url#1{\texttt{#1}}\fi
\expandafter\ifx\csname urlprefix\endcsname\relax\def\urlprefix{URL }\fi
\providecommand{\bibinfo}[2]{#2}
\providecommand{\eprint}[2][]{\url{#2}}

\bibitem[{\citenamefont{Gr{\"u}nberg et~al.}(1986)\citenamefont{Gr{\"u}nberg,
  Schreiber, Pang, Brodsky, and Sowers}}]{grunberg1986}
\bibinfo{author}{\bibfnamefont{P.}~\bibnamefont{Gr{\"u}nberg}},
  \bibinfo{author}{\bibfnamefont{R.}~\bibnamefont{Schreiber}},
  \bibinfo{author}{\bibfnamefont{Y.}~\bibnamefont{Pang}},
  \bibinfo{author}{\bibfnamefont{M.}~\bibnamefont{Brodsky}}, \bibnamefont{and}
  \bibinfo{author}{\bibfnamefont{H.}~\bibnamefont{Sowers}},
  \bibinfo{journal}{Physical Review Letters} \textbf{\bibinfo{volume}{57}},
  \bibinfo{pages}{2442} (\bibinfo{year}{1986}).

\bibitem[{\citenamefont{Baibich et~al.}(1988)\citenamefont{Baibich, Broto,
  Fert, Van~Dau, Petroff, Etienne, Creuzet, Friederich, and
  Chazelas}}]{baibich1988}
\bibinfo{author}{\bibfnamefont{M.~N.} \bibnamefont{Baibich}},
  \bibinfo{author}{\bibfnamefont{J.~M.} \bibnamefont{Broto}},
  \bibinfo{author}{\bibfnamefont{A.}~\bibnamefont{Fert}},
  \bibinfo{author}{\bibfnamefont{F.~N.} \bibnamefont{Van~Dau}},
  \bibinfo{author}{\bibfnamefont{F.}~\bibnamefont{Petroff}},
  \bibinfo{author}{\bibfnamefont{P.}~\bibnamefont{Etienne}},
  \bibinfo{author}{\bibfnamefont{G.}~\bibnamefont{Creuzet}},
  \bibinfo{author}{\bibfnamefont{A.}~\bibnamefont{Friederich}},
  \bibnamefont{and} \bibinfo{author}{\bibfnamefont{J.}~\bibnamefont{Chazelas}},
  \bibinfo{journal}{Physical Review Letters} \textbf{\bibinfo{volume}{61}},
  \bibinfo{pages}{2472} (\bibinfo{year}{1988}).

\bibitem[{\citenamefont{Xiao et~al.}(1992)\citenamefont{Xiao, Jiang, and
  Chien}}]{xiao1992}
\bibinfo{author}{\bibfnamefont{J.~Q.} \bibnamefont{Xiao}},
  \bibinfo{author}{\bibfnamefont{J.~S.} \bibnamefont{Jiang}}, \bibnamefont{and}
  \bibinfo{author}{\bibfnamefont{C.}~\bibnamefont{Chien}},
  \bibinfo{journal}{Physical Review Letters} \textbf{\bibinfo{volume}{68}},
  \bibinfo{pages}{3749} (\bibinfo{year}{1992}).

\bibitem[{\citenamefont{Thompson}(2008)}]{thompson2008}
\bibinfo{author}{\bibfnamefont{S.~M.} \bibnamefont{Thompson}},
  \bibinfo{journal}{Journal of Physics D: Applied Physics}
  \textbf{\bibinfo{volume}{41}}, \bibinfo{pages}{093001}
  (\bibinfo{year}{2008}).

\bibitem[{\citenamefont{Ennen et~al.}(2016)\citenamefont{Ennen, Kappe, Rempel,
  Glenske, and H{\"u}tten}}]{ennen2016}
\bibinfo{author}{\bibfnamefont{I.}~\bibnamefont{Ennen}},
  \bibinfo{author}{\bibfnamefont{D.}~\bibnamefont{Kappe}},
  \bibinfo{author}{\bibfnamefont{T.}~\bibnamefont{Rempel}},
  \bibinfo{author}{\bibfnamefont{C.}~\bibnamefont{Glenske}}, \bibnamefont{and}
  \bibinfo{author}{\bibfnamefont{A.}~\bibnamefont{H{\"u}tten}},
  \bibinfo{journal}{Sensors} \textbf{\bibinfo{volume}{16}},
  \bibinfo{pages}{904} (\bibinfo{year}{2016}).

\bibitem[{\citenamefont{Di{\'e}ny}(1994)}]{dieny1994}
\bibinfo{author}{\bibfnamefont{B.}~\bibnamefont{Di{\'e}ny}},
  \bibinfo{journal}{Journal of Magnetism and Magnetic Materials}
  \textbf{\bibinfo{volume}{136}}, \bibinfo{pages}{335} (\bibinfo{year}{1994}).

\bibitem[{\citenamefont{von Helmolt et~al.}(1993)\citenamefont{von Helmolt,
  Wecker, Holzapfel, Schultz, and Samwer}}]{von1993}
\bibinfo{author}{\bibfnamefont{R.}~\bibnamefont{von Helmolt}},
  \bibinfo{author}{\bibfnamefont{J.}~\bibnamefont{Wecker}},
  \bibinfo{author}{\bibfnamefont{B.}~\bibnamefont{Holzapfel}},
  \bibinfo{author}{\bibfnamefont{L.}~\bibnamefont{Schultz}}, \bibnamefont{and}
  \bibinfo{author}{\bibfnamefont{K.}~\bibnamefont{Samwer}},
  \bibinfo{journal}{Physical Review Letters} \textbf{\bibinfo{volume}{71}},
  \bibinfo{pages}{2331} (\bibinfo{year}{1993}).

\bibitem[{\citenamefont{Jin et~al.}(1994)\citenamefont{Jin, Tiefel, McCormack,
  Fastnacht, Ramesh, and Chen}}]{jin1994}
\bibinfo{author}{\bibfnamefont{S.}~\bibnamefont{Jin}},
  \bibinfo{author}{\bibfnamefont{T.~H.} \bibnamefont{Tiefel}},
  \bibinfo{author}{\bibfnamefont{M.}~\bibnamefont{McCormack}},
  \bibinfo{author}{\bibfnamefont{R.}~\bibnamefont{Fastnacht}},
  \bibinfo{author}{\bibfnamefont{R.}~\bibnamefont{Ramesh}}, \bibnamefont{and}
  \bibinfo{author}{\bibfnamefont{L.}~\bibnamefont{Chen}},
  \bibinfo{journal}{Science} \textbf{\bibinfo{volume}{264}},
  \bibinfo{pages}{413} (\bibinfo{year}{1994}).

\bibitem[{\citenamefont{Millis et~al.}(1996)\citenamefont{Millis, Shraiman, and
  Mueller}}]{millis1996}
\bibinfo{author}{\bibfnamefont{A.}~\bibnamefont{Millis}},
  \bibinfo{author}{\bibfnamefont{B.~I.} \bibnamefont{Shraiman}},
  \bibnamefont{and} \bibinfo{author}{\bibfnamefont{R.}~\bibnamefont{Mueller}},
  \bibinfo{journal}{Physical Review Letters} \textbf{\bibinfo{volume}{77}},
  \bibinfo{pages}{175} (\bibinfo{year}{1996}).

\bibitem[{\citenamefont{Wang et~al.}(2021)\citenamefont{Wang, Rogers, Yao,
  Nichols, Atay, Xu, Franklin, Sochnikov, Ryan, Haskel et~al.}}]{wang2021}
\bibinfo{author}{\bibfnamefont{Z.-C.} \bibnamefont{Wang}},
  \bibinfo{author}{\bibfnamefont{J.~D.} \bibnamefont{Rogers}},
  \bibinfo{author}{\bibfnamefont{X.}~\bibnamefont{Yao}},
  \bibinfo{author}{\bibfnamefont{R.}~\bibnamefont{Nichols}},
  \bibinfo{author}{\bibfnamefont{K.}~\bibnamefont{Atay}},
  \bibinfo{author}{\bibfnamefont{B.}~\bibnamefont{Xu}},
  \bibinfo{author}{\bibfnamefont{J.}~\bibnamefont{Franklin}},
  \bibinfo{author}{\bibfnamefont{I.}~\bibnamefont{Sochnikov}},
  \bibinfo{author}{\bibfnamefont{P.~J.} \bibnamefont{Ryan}},
  \bibinfo{author}{\bibfnamefont{D.}~\bibnamefont{Haskel}},
  \bibnamefont{et~al.}, \bibinfo{journal}{Advanced Materials}
  \textbf{\bibinfo{volume}{33}}, \bibinfo{pages}{2005755}
  (\bibinfo{year}{2021}).

\bibitem[{\citenamefont{Shimakawa et~al.}(1996)\citenamefont{Shimakawa, Kubo,
  and Manako}}]{shimakawa1996}
\bibinfo{author}{\bibfnamefont{Y.}~\bibnamefont{Shimakawa}},
  \bibinfo{author}{\bibfnamefont{Y.}~\bibnamefont{Kubo}}, \bibnamefont{and}
  \bibinfo{author}{\bibfnamefont{T.}~\bibnamefont{Manako}},
  \bibinfo{journal}{Nature} \textbf{\bibinfo{volume}{379}}, \bibinfo{pages}{53}
  (\bibinfo{year}{1996}).

\bibitem[{\citenamefont{Majumdar and Littlewood}(1998)}]{majumdar1998}
\bibinfo{author}{\bibfnamefont{P.}~\bibnamefont{Majumdar}} \bibnamefont{and}
  \bibinfo{author}{\bibfnamefont{P.}~\bibnamefont{Littlewood}},
  \bibinfo{journal}{Physical Review Letters} \textbf{\bibinfo{volume}{81}},
  \bibinfo{pages}{1314} (\bibinfo{year}{1998}).

\bibitem[{\citenamefont{Rosa et~al.}(2020)\citenamefont{Rosa, Xu, Rahn, Souza,
  Kushwaha, Veiga, Bombardi, Thomas, Janoschek, Bauer et~al.}}]{rosa2020}
\bibinfo{author}{\bibfnamefont{P.}~\bibnamefont{Rosa}},
  \bibinfo{author}{\bibfnamefont{Y.}~\bibnamefont{Xu}},
  \bibinfo{author}{\bibfnamefont{M.}~\bibnamefont{Rahn}},
  \bibinfo{author}{\bibfnamefont{J.}~\bibnamefont{Souza}},
  \bibinfo{author}{\bibfnamefont{S.}~\bibnamefont{Kushwaha}},
  \bibinfo{author}{\bibfnamefont{L.}~\bibnamefont{Veiga}},
  \bibinfo{author}{\bibfnamefont{A.}~\bibnamefont{Bombardi}},
  \bibinfo{author}{\bibfnamefont{S.}~\bibnamefont{Thomas}},
  \bibinfo{author}{\bibfnamefont{M.}~\bibnamefont{Janoschek}},
  \bibinfo{author}{\bibfnamefont{E.}~\bibnamefont{Bauer}},
  \bibnamefont{et~al.}, \bibinfo{journal}{npj Quantum Materials}
  \textbf{\bibinfo{volume}{5}}, \bibinfo{pages}{52} (\bibinfo{year}{2020}).

\bibitem[{\citenamefont{Yin et~al.}(2020)\citenamefont{Yin, Wu, Li, Yu, Sun,
  Shen, Frandsen, Yao, and Wang}}]{yin2020}
\bibinfo{author}{\bibfnamefont{J.}~\bibnamefont{Yin}},
  \bibinfo{author}{\bibfnamefont{C.}~\bibnamefont{Wu}},
  \bibinfo{author}{\bibfnamefont{L.}~\bibnamefont{Li}},
  \bibinfo{author}{\bibfnamefont{J.}~\bibnamefont{Yu}},
  \bibinfo{author}{\bibfnamefont{H.}~\bibnamefont{Sun}},
  \bibinfo{author}{\bibfnamefont{B.}~\bibnamefont{Shen}},
  \bibinfo{author}{\bibfnamefont{B.~A.} \bibnamefont{Frandsen}},
  \bibinfo{author}{\bibfnamefont{D.-X.} \bibnamefont{Yao}}, \bibnamefont{and}
  \bibinfo{author}{\bibfnamefont{M.}~\bibnamefont{Wang}},
  \bibinfo{journal}{Physical Review Materials} \textbf{\bibinfo{volume}{4}},
  \bibinfo{pages}{013405} (\bibinfo{year}{2020}).

\bibitem[{\citenamefont{Yang et~al.}(2021)\citenamefont{Yang, Liu, Liao, Si,
  Jiang, Liu, Guo, Yin, Wang, Sheng et~al.}}]{yang2021}
\bibinfo{author}{\bibfnamefont{H.}~\bibnamefont{Yang}},
  \bibinfo{author}{\bibfnamefont{Q.}~\bibnamefont{Liu}},
  \bibinfo{author}{\bibfnamefont{Z.}~\bibnamefont{Liao}},
  \bibinfo{author}{\bibfnamefont{L.}~\bibnamefont{Si}},
  \bibinfo{author}{\bibfnamefont{P.}~\bibnamefont{Jiang}},
  \bibinfo{author}{\bibfnamefont{X.}~\bibnamefont{Liu}},
  \bibinfo{author}{\bibfnamefont{Y.}~\bibnamefont{Guo}},
  \bibinfo{author}{\bibfnamefont{J.}~\bibnamefont{Yin}},
  \bibinfo{author}{\bibfnamefont{M.}~\bibnamefont{Wang}},
  \bibinfo{author}{\bibfnamefont{Z.}~\bibnamefont{Sheng}},
  \bibnamefont{et~al.}, \bibinfo{journal}{Physical Review B}
  \textbf{\bibinfo{volume}{104}}, \bibinfo{pages}{214419}
  (\bibinfo{year}{2021}).

\bibitem[{\citenamefont{Sun et~al.}(2021{\natexlab{a}})\citenamefont{Sun, Wang,
  Mu, Wang, Wang, Wu, Wang, Zhou, and Chen}}]{sun2021}
\bibinfo{author}{\bibfnamefont{Z.}~\bibnamefont{Sun}},
  \bibinfo{author}{\bibfnamefont{A.}~\bibnamefont{Wang}},
  \bibinfo{author}{\bibfnamefont{H.}~\bibnamefont{Mu}},
  \bibinfo{author}{\bibfnamefont{H.}~\bibnamefont{Wang}},
  \bibinfo{author}{\bibfnamefont{Z.}~\bibnamefont{Wang}},
  \bibinfo{author}{\bibfnamefont{T.}~\bibnamefont{Wu}},
  \bibinfo{author}{\bibfnamefont{Z.}~\bibnamefont{Wang}},
  \bibinfo{author}{\bibfnamefont{X.}~\bibnamefont{Zhou}}, \bibnamefont{and}
  \bibinfo{author}{\bibfnamefont{X.}~\bibnamefont{Chen}}, \bibinfo{journal}{npj
  Quantum Materials} \textbf{\bibinfo{volume}{6}}, \bibinfo{pages}{94}
  (\bibinfo{year}{2021}{\natexlab{a}}).

\bibitem[{\citenamefont{Ogasawara et~al.}(2021)\citenamefont{Ogasawara, Huynh,
  Tahara, Kida, Hagiwara, Ar{\v{c}}on, Kimata, Matsushita, Nagata, and
  Tanigaki}}]{ogasawara2021}
\bibinfo{author}{\bibfnamefont{T.}~\bibnamefont{Ogasawara}},
  \bibinfo{author}{\bibfnamefont{K.-K.} \bibnamefont{Huynh}},
  \bibinfo{author}{\bibfnamefont{T.}~\bibnamefont{Tahara}},
  \bibinfo{author}{\bibfnamefont{T.}~\bibnamefont{Kida}},
  \bibinfo{author}{\bibfnamefont{M.}~\bibnamefont{Hagiwara}},
  \bibinfo{author}{\bibfnamefont{D.}~\bibnamefont{Ar{\v{c}}on}},
  \bibinfo{author}{\bibfnamefont{M.}~\bibnamefont{Kimata}},
  \bibinfo{author}{\bibfnamefont{S.~Y.} \bibnamefont{Matsushita}},
  \bibinfo{author}{\bibfnamefont{K.}~\bibnamefont{Nagata}}, \bibnamefont{and}
  \bibinfo{author}{\bibfnamefont{K.}~\bibnamefont{Tanigaki}},
  \bibinfo{journal}{Physical Review B} \textbf{\bibinfo{volume}{103}},
  \bibinfo{pages}{125108} (\bibinfo{year}{2021}).

\bibitem[{\citenamefont{Vincent et~al.}(1986)\citenamefont{Vincent, Leroux,
  Bijaoui, Rimet, and Schlenker}}]{vincent1986}
\bibinfo{author}{\bibfnamefont{H.}~\bibnamefont{Vincent}},
  \bibinfo{author}{\bibfnamefont{D.}~\bibnamefont{Leroux}},
  \bibinfo{author}{\bibfnamefont{D.}~\bibnamefont{Bijaoui}},
  \bibinfo{author}{\bibfnamefont{R.}~\bibnamefont{Rimet}}, \bibnamefont{and}
  \bibinfo{author}{\bibfnamefont{C.}~\bibnamefont{Schlenker}},
  \bibinfo{journal}{Journal of Solid State Chemistry}
  \textbf{\bibinfo{volume}{63}}, \bibinfo{pages}{349} (\bibinfo{year}{1986}).

\bibitem[{\citenamefont{Ni et~al.}(2021)\citenamefont{Ni, Zhao, Zhang, Hu,
  Kimchi, and Cao}}]{ni2021}
\bibinfo{author}{\bibfnamefont{Y.}~\bibnamefont{Ni}},
  \bibinfo{author}{\bibfnamefont{H.}~\bibnamefont{Zhao}},
  \bibinfo{author}{\bibfnamefont{Y.}~\bibnamefont{Zhang}},
  \bibinfo{author}{\bibfnamefont{B.}~\bibnamefont{Hu}},
  \bibinfo{author}{\bibfnamefont{I.}~\bibnamefont{Kimchi}}, \bibnamefont{and}
  \bibinfo{author}{\bibfnamefont{G.}~\bibnamefont{Cao}},
  \bibinfo{journal}{Physical Review B} \textbf{\bibinfo{volume}{103}},
  \bibinfo{pages}{L161105} (\bibinfo{year}{2021}).

\bibitem[{\citenamefont{Seo et~al.}(2021)\citenamefont{Seo, De, Ha, Lee, Park,
  Park, Skourski, Choi, Kim, Cho et~al.}}]{seo2021}
\bibinfo{author}{\bibfnamefont{J.}~\bibnamefont{Seo}},
  \bibinfo{author}{\bibfnamefont{C.}~\bibnamefont{De}},
  \bibinfo{author}{\bibfnamefont{H.}~\bibnamefont{Ha}},
  \bibinfo{author}{\bibfnamefont{J.~E.} \bibnamefont{Lee}},
  \bibinfo{author}{\bibfnamefont{S.}~\bibnamefont{Park}},
  \bibinfo{author}{\bibfnamefont{J.}~\bibnamefont{Park}},
  \bibinfo{author}{\bibfnamefont{Y.}~\bibnamefont{Skourski}},
  \bibinfo{author}{\bibfnamefont{E.~S.} \bibnamefont{Choi}},
  \bibinfo{author}{\bibfnamefont{B.}~\bibnamefont{Kim}},
  \bibinfo{author}{\bibfnamefont{G.~Y.} \bibnamefont{Cho}},
  \bibnamefont{et~al.}, \bibinfo{journal}{Nature}
  \textbf{\bibinfo{volume}{599}}, \bibinfo{pages}{576} (\bibinfo{year}{2021}).

\bibitem[{\citenamefont{Sala et~al.}(2022)\citenamefont{Sala, Lin, Samarakoon,
  Parker, May, and Stone}}]{sala2022}
\bibinfo{author}{\bibfnamefont{G.}~\bibnamefont{Sala}},
  \bibinfo{author}{\bibfnamefont{J.}~\bibnamefont{Lin}},
  \bibinfo{author}{\bibfnamefont{A.}~\bibnamefont{Samarakoon}},
  \bibinfo{author}{\bibfnamefont{D.}~\bibnamefont{Parker}},
  \bibinfo{author}{\bibfnamefont{A.}~\bibnamefont{May}}, \bibnamefont{and}
  \bibinfo{author}{\bibfnamefont{M.}~\bibnamefont{Stone}},
  \bibinfo{journal}{Physical Review B} \textbf{\bibinfo{volume}{105}},
  \bibinfo{pages}{214405} (\bibinfo{year}{2022}).

\bibitem[{\citenamefont{Kwon et~al.}(2023)\citenamefont{Kwon, Kim, Kim, Susilo,
  Kang, Kim, Kim, Kim, Kim, and Kim}}]{kwon2023}
\bibinfo{author}{\bibfnamefont{C.~I.} \bibnamefont{Kwon}},
  \bibinfo{author}{\bibfnamefont{K.}~\bibnamefont{Kim}},
  \bibinfo{author}{\bibfnamefont{S.~Y.} \bibnamefont{Kim}},
  \bibinfo{author}{\bibfnamefont{R.~A.} \bibnamefont{Susilo}},
  \bibinfo{author}{\bibfnamefont{B.}~\bibnamefont{Kang}},
  \bibinfo{author}{\bibfnamefont{K.}~\bibnamefont{Kim}},
  \bibinfo{author}{\bibfnamefont{D.~Y.} \bibnamefont{Kim}},
  \bibinfo{author}{\bibfnamefont{J.}~\bibnamefont{Kim}},
  \bibinfo{author}{\bibfnamefont{B.}~\bibnamefont{Kim}}, \bibnamefont{and}
  \bibinfo{author}{\bibfnamefont{J.~S.} \bibnamefont{Kim}},
  \bibinfo{journal}{Current Applied Physics} \textbf{\bibinfo{volume}{53}},
  \bibinfo{pages}{51} (\bibinfo{year}{2023}).

\bibitem[{\citenamefont{Mijin et~al.}(2023)\citenamefont{Mijin,
  {\v{S}}olaji{\'c}, Pe{\v{s}}i{\'c}, Liu, Petrovic, Bockstedte, Bonanni,
  Popovi{\'c}, and Lazarevi{\'c}}}]{mijin2023}
\bibinfo{author}{\bibfnamefont{S.~D.} \bibnamefont{Mijin}},
  \bibinfo{author}{\bibfnamefont{A.}~\bibnamefont{{\v{S}}olaji{\'c}}},
  \bibinfo{author}{\bibfnamefont{J.}~\bibnamefont{Pe{\v{s}}i{\'c}}},
  \bibinfo{author}{\bibfnamefont{Y.}~\bibnamefont{Liu}},
  \bibinfo{author}{\bibfnamefont{C.}~\bibnamefont{Petrovic}},
  \bibinfo{author}{\bibfnamefont{M.}~\bibnamefont{Bockstedte}},
  \bibinfo{author}{\bibfnamefont{A.}~\bibnamefont{Bonanni}},
  \bibinfo{author}{\bibfnamefont{Z.}~\bibnamefont{Popovi{\'c}}},
  \bibnamefont{and}
  \bibinfo{author}{\bibfnamefont{N.}~\bibnamefont{Lazarevi{\'c}}},
  \bibinfo{journal}{Physical Review B} \textbf{\bibinfo{volume}{107}},
  \bibinfo{pages}{054309} (\bibinfo{year}{2023}).

\bibitem[{\citenamefont{Lovesey}(2023)}]{lovesey2023}
\bibinfo{author}{\bibfnamefont{S.~W.} \bibnamefont{Lovesey}},
  \bibinfo{journal}{Physical Review B} \textbf{\bibinfo{volume}{107}},
  \bibinfo{pages}{224410} (\bibinfo{year}{2023}).

\bibitem[{\citenamefont{Zhang et~al.}(2022{\natexlab{a}})\citenamefont{Zhang,
  Ni, Zhao, Hakani, Ye, DeLong, Kimchi, and Cao}}]{zhang2022Control}
\bibinfo{author}{\bibfnamefont{Y.}~\bibnamefont{Zhang}},
  \bibinfo{author}{\bibfnamefont{Y.}~\bibnamefont{Ni}},
  \bibinfo{author}{\bibfnamefont{H.}~\bibnamefont{Zhao}},
  \bibinfo{author}{\bibfnamefont{S.}~\bibnamefont{Hakani}},
  \bibinfo{author}{\bibfnamefont{F.}~\bibnamefont{Ye}},
  \bibinfo{author}{\bibfnamefont{L.}~\bibnamefont{DeLong}},
  \bibinfo{author}{\bibfnamefont{I.}~\bibnamefont{Kimchi}}, \bibnamefont{and}
  \bibinfo{author}{\bibfnamefont{G.}~\bibnamefont{Cao}},
  \bibinfo{journal}{Nature} \textbf{\bibinfo{volume}{611}},
  \bibinfo{pages}{467} (\bibinfo{year}{2022}{\natexlab{a}}).

\bibitem[{\citenamefont{Ye et~al.}(2022)\citenamefont{Ye, Matsuda, Morgan,
  Sherline, Ni, Zhao, and Cao}}]{ye2022}
\bibinfo{author}{\bibfnamefont{F.}~\bibnamefont{Ye}},
  \bibinfo{author}{\bibfnamefont{M.}~\bibnamefont{Matsuda}},
  \bibinfo{author}{\bibfnamefont{Z.}~\bibnamefont{Morgan}},
  \bibinfo{author}{\bibfnamefont{T.}~\bibnamefont{Sherline}},
  \bibinfo{author}{\bibfnamefont{Y.}~\bibnamefont{Ni}},
  \bibinfo{author}{\bibfnamefont{H.}~\bibnamefont{Zhao}}, \bibnamefont{and}
  \bibinfo{author}{\bibfnamefont{G.}~\bibnamefont{Cao}},
  \bibinfo{journal}{Physical Review B} \textbf{\bibinfo{volume}{106}},
  \bibinfo{pages}{L180402} (\bibinfo{year}{2022}).

\bibitem[{\citenamefont{Zhang et~al.}(2022{\natexlab{b}})\citenamefont{Zhang,
  Liu, Cao, Phelan, Graf, DiTusa, Tennant, and Mao}}]{zhang2022}
\bibinfo{author}{\bibfnamefont{Q.}~\bibnamefont{Zhang}},
  \bibinfo{author}{\bibfnamefont{J.}~\bibnamefont{Liu}},
  \bibinfo{author}{\bibfnamefont{H.}~\bibnamefont{Cao}},
  \bibinfo{author}{\bibfnamefont{A.}~\bibnamefont{Phelan}},
  \bibinfo{author}{\bibfnamefont{D.}~\bibnamefont{Graf}},
  \bibinfo{author}{\bibfnamefont{J.}~\bibnamefont{DiTusa}},
  \bibinfo{author}{\bibfnamefont{D.~A.} \bibnamefont{Tennant}},
  \bibnamefont{and} \bibinfo{author}{\bibfnamefont{Z.}~\bibnamefont{Mao}},
  \bibinfo{journal}{NPG Asia Materials} \textbf{\bibinfo{volume}{14}},
  \bibinfo{pages}{1884} (\bibinfo{year}{2022}{\natexlab{b}}).

\bibitem[{\citenamefont{Zhang et~al.}(2023)\citenamefont{Zhang, Lin, Moreo, and
  Dagotto}}]{zhang2023}
\bibinfo{author}{\bibfnamefont{Y.}~\bibnamefont{Zhang}},
  \bibinfo{author}{\bibfnamefont{L.-F.} \bibnamefont{Lin}},
  \bibinfo{author}{\bibfnamefont{A.}~\bibnamefont{Moreo}}, \bibnamefont{and}
  \bibinfo{author}{\bibfnamefont{E.}~\bibnamefont{Dagotto}},
  \bibinfo{journal}{Physical Review B} \textbf{\bibinfo{volume}{107}},
  \bibinfo{pages}{054430} (\bibinfo{year}{2023}).

\bibitem[{\citenamefont{Sun et~al.}(2023{\natexlab{a}})\citenamefont{Sun, Qiu,
  Han, Yi, Li, Huo, Huang, Liu, Li, Wang et~al.}}]{sun2023coexistence}
\bibinfo{author}{\bibfnamefont{H.}~\bibnamefont{Sun}},
  \bibinfo{author}{\bibfnamefont{L.}~\bibnamefont{Qiu}},
  \bibinfo{author}{\bibfnamefont{Y.}~\bibnamefont{Han}},
  \bibinfo{author}{\bibfnamefont{E.}~\bibnamefont{Yi}},
  \bibinfo{author}{\bibfnamefont{J.}~\bibnamefont{Li}},
  \bibinfo{author}{\bibfnamefont{M.}~\bibnamefont{Huo}},
  \bibinfo{author}{\bibfnamefont{C.}~\bibnamefont{Huang}},
  \bibinfo{author}{\bibfnamefont{H.}~\bibnamefont{Liu}},
  \bibinfo{author}{\bibfnamefont{M.}~\bibnamefont{Li}},
  \bibinfo{author}{\bibfnamefont{W.}~\bibnamefont{Wang}}, \bibnamefont{et~al.},
  \bibinfo{journal}{Materials Today Physics} \textbf{\bibinfo{volume}{36}},
  \bibinfo{pages}{101188} (\bibinfo{year}{2023}{\natexlab{a}}).

\bibitem[{\citenamefont{Sun et~al.}(2023{\natexlab{b}})\citenamefont{Sun, Qiu,
  Han, Zhang, Wang, Huang, Liu, Huo, Li, Liu et~al.}}]{sun2023exchange}
\bibinfo{author}{\bibfnamefont{H.}~\bibnamefont{Sun}},
  \bibinfo{author}{\bibfnamefont{L.}~\bibnamefont{Qiu}},
  \bibinfo{author}{\bibfnamefont{Y.}~\bibnamefont{Han}},
  \bibinfo{author}{\bibfnamefont{Y.}~\bibnamefont{Zhang}},
  \bibinfo{author}{\bibfnamefont{W.}~\bibnamefont{Wang}},
  \bibinfo{author}{\bibfnamefont{C.}~\bibnamefont{Huang}},
  \bibinfo{author}{\bibfnamefont{N.}~\bibnamefont{Liu}},
  \bibinfo{author}{\bibfnamefont{M.}~\bibnamefont{Huo}},
  \bibinfo{author}{\bibfnamefont{L.}~\bibnamefont{Li}},
  \bibinfo{author}{\bibfnamefont{H.}~\bibnamefont{Liu}}, \bibnamefont{et~al.},
  \bibinfo{journal}{Communications Physics} \textbf{\bibinfo{volume}{6}},
  \bibinfo{pages}{40} (\bibinfo{year}{2023}{\natexlab{b}}).

\bibitem[{\citenamefont{Sun et~al.}(2021{\natexlab{b}})\citenamefont{Sun, Chen,
  Hou, Wang, Gong, Huo, Li, Yu, Cai, Liu et~al.}}]{sun2021magnetism}
\bibinfo{author}{\bibfnamefont{H.}~\bibnamefont{Sun}},
  \bibinfo{author}{\bibfnamefont{C.}~\bibnamefont{Chen}},
  \bibinfo{author}{\bibfnamefont{Y.}~\bibnamefont{Hou}},
  \bibinfo{author}{\bibfnamefont{W.}~\bibnamefont{Wang}},
  \bibinfo{author}{\bibfnamefont{Y.}~\bibnamefont{Gong}},
  \bibinfo{author}{\bibfnamefont{M.}~\bibnamefont{Huo}},
  \bibinfo{author}{\bibfnamefont{L.}~\bibnamefont{Li}},
  \bibinfo{author}{\bibfnamefont{J.}~\bibnamefont{Yu}},
  \bibinfo{author}{\bibfnamefont{W.}~\bibnamefont{Cai}},
  \bibinfo{author}{\bibfnamefont{N.}~\bibnamefont{Liu}}, \bibnamefont{et~al.},
  \bibinfo{journal}{Science China Physics, Mechanics \& Astronomy}
  \textbf{\bibinfo{volume}{64}}, \bibinfo{pages}{118211}
  (\bibinfo{year}{2021}{\natexlab{b}}).

\bibitem[{\citenamefont{Cai et~al.}(2020)\citenamefont{Cai, Sun, Xia, Wu, Liu,
  Liu, Gong, Yao, Guo, and Wang}}]{Cai2020}
\bibinfo{author}{\bibfnamefont{W.}~\bibnamefont{Cai}},
  \bibinfo{author}{\bibfnamefont{H.}~\bibnamefont{Sun}},
  \bibinfo{author}{\bibfnamefont{W.}~\bibnamefont{Xia}},
  \bibinfo{author}{\bibfnamefont{C.}~\bibnamefont{Wu}},
  \bibinfo{author}{\bibfnamefont{Y.}~\bibnamefont{Liu}},
  \bibinfo{author}{\bibfnamefont{H.}~\bibnamefont{Liu}},
  \bibinfo{author}{\bibfnamefont{Y.}~\bibnamefont{Gong}},
  \bibinfo{author}{\bibfnamefont{D.-X.} \bibnamefont{Yao}},
  \bibinfo{author}{\bibfnamefont{Y.}~\bibnamefont{Guo}}, \bibnamefont{and}
  \bibinfo{author}{\bibfnamefont{M.}~\bibnamefont{Wang}},
  \bibinfo{journal}{Physical Review B} \textbf{\bibinfo{volume}{102}},
  \bibinfo{pages}{144525} (\bibinfo{year}{2020}).

\bibitem[{\citenamefont{Prescher and Prakapenka}(2015)}]{Clemens2015}
\bibinfo{author}{\bibfnamefont{C.}~\bibnamefont{Prescher}} \bibnamefont{and}
  \bibinfo{author}{\bibfnamefont{V.~B.} \bibnamefont{Prakapenka}},
  \bibinfo{journal}{High Pressure Research} \textbf{\bibinfo{volume}{35}},
  \bibinfo{pages}{233} (\bibinfo{year}{2015}).

\bibitem[{\citenamefont{Pawley}(1981)}]{Pawley1981}
\bibinfo{author}{\bibfnamefont{G.~S.} \bibnamefont{Pawley}},
  \bibinfo{journal}{Journal of Applied Crystallography}
  \textbf{\bibinfo{volume}{14}}, \bibinfo{pages}{357} (\bibinfo{year}{1981}),
  ISSN \bibinfo{issn}{0021-8898}.

\bibitem[{\citenamefont{Xu et~al.}(2018)\citenamefont{Xu, Li, Wang, Chen, Wu,
  Zhang, Li, Lin, Chen, and Pei}}]{xu2018}
\bibinfo{author}{\bibfnamefont{Y.}~\bibnamefont{Xu}},
  \bibinfo{author}{\bibfnamefont{W.}~\bibnamefont{Li}},
  \bibinfo{author}{\bibfnamefont{C.}~\bibnamefont{Wang}},
  \bibinfo{author}{\bibfnamefont{Z.}~\bibnamefont{Chen}},
  \bibinfo{author}{\bibfnamefont{Y.}~\bibnamefont{Wu}},
  \bibinfo{author}{\bibfnamefont{X.}~\bibnamefont{Zhang}},
  \bibinfo{author}{\bibfnamefont{J.}~\bibnamefont{Li}},
  \bibinfo{author}{\bibfnamefont{S.}~\bibnamefont{Lin}},
  \bibinfo{author}{\bibfnamefont{Y.}~\bibnamefont{Chen}}, \bibnamefont{and}
  \bibinfo{author}{\bibfnamefont{Y.}~\bibnamefont{Pei}},
  \bibinfo{journal}{Journal of Materiomics} \textbf{\bibinfo{volume}{4}},
  \bibinfo{pages}{215} (\bibinfo{year}{2018}).

\bibitem[{\citenamefont{May et~al.}(2017)\citenamefont{May, Liu, Calder,
  Parker, Pandey, Cakmak, Cao, Yan, and McGuire}}]{may2017}
\bibinfo{author}{\bibfnamefont{A.~F.} \bibnamefont{May}},
  \bibinfo{author}{\bibfnamefont{Y.}~\bibnamefont{Liu}},
  \bibinfo{author}{\bibfnamefont{S.}~\bibnamefont{Calder}},
  \bibinfo{author}{\bibfnamefont{D.~S.} \bibnamefont{Parker}},
  \bibinfo{author}{\bibfnamefont{T.}~\bibnamefont{Pandey}},
  \bibinfo{author}{\bibfnamefont{E.}~\bibnamefont{Cakmak}},
  \bibinfo{author}{\bibfnamefont{H.}~\bibnamefont{Cao}},
  \bibinfo{author}{\bibfnamefont{J.}~\bibnamefont{Yan}}, \bibnamefont{and}
  \bibinfo{author}{\bibfnamefont{M.~A.} \bibnamefont{McGuire}},
  \bibinfo{journal}{Physical Review B} \textbf{\bibinfo{volume}{95}},
  \bibinfo{pages}{174440} (\bibinfo{year}{2017}).

\bibitem[{\citenamefont{Liu et~al.}(2018)\citenamefont{Liu, Petrovic
  et~al.}}]{liu2018}
\bibinfo{author}{\bibfnamefont{Y.}~\bibnamefont{Liu}},
  \bibinfo{author}{\bibfnamefont{C.}~\bibnamefont{Petrovic}},
  \bibnamefont{et~al.}, \bibinfo{journal}{Physical Review B}
  \textbf{\bibinfo{volume}{98}}, \bibinfo{pages}{064423}
  (\bibinfo{year}{2018}).

\bibitem[{\citenamefont{Martinez et~al.}(2020)\citenamefont{Martinez,
  Iturriaga, Olmos, Shao, Liu, Mai, Petrovic, Hight~Walker, and
  Singamaneni}}]{martinez2020}
\bibinfo{author}{\bibfnamefont{L.}~\bibnamefont{Martinez}},
  \bibinfo{author}{\bibfnamefont{H.}~\bibnamefont{Iturriaga}},
  \bibinfo{author}{\bibfnamefont{R.}~\bibnamefont{Olmos}},
  \bibinfo{author}{\bibfnamefont{L.}~\bibnamefont{Shao}},
  \bibinfo{author}{\bibfnamefont{Y.}~\bibnamefont{Liu}},
  \bibinfo{author}{\bibfnamefont{T.~T.} \bibnamefont{Mai}},
  \bibinfo{author}{\bibfnamefont{C.}~\bibnamefont{Petrovic}},
  \bibinfo{author}{\bibfnamefont{A.~R.} \bibnamefont{Hight~Walker}},
  \bibnamefont{and}
  \bibinfo{author}{\bibfnamefont{S.}~\bibnamefont{Singamaneni}},
  \bibinfo{journal}{Applied Physics Letters} \textbf{\bibinfo{volume}{116}},
  \bibinfo{pages}{172404} (\bibinfo{year}{2020}).

\bibitem[{\citenamefont{Liu et~al.}(2021)\citenamefont{Liu, Hu, Abeykoon,
  Stavitski, Attenkofer, Bauer, Petrovic et~al.}}]{liu2021polaronic}
\bibinfo{author}{\bibfnamefont{Y.}~\bibnamefont{Liu}},
  \bibinfo{author}{\bibfnamefont{Z.}~\bibnamefont{Hu}},
  \bibinfo{author}{\bibfnamefont{M.}~\bibnamefont{Abeykoon}},
  \bibinfo{author}{\bibfnamefont{E.}~\bibnamefont{Stavitski}},
  \bibinfo{author}{\bibfnamefont{K.}~\bibnamefont{Attenkofer}},
  \bibinfo{author}{\bibfnamefont{E.~D.} \bibnamefont{Bauer}},
  \bibinfo{author}{\bibfnamefont{C.}~\bibnamefont{Petrovic}},
  \bibnamefont{et~al.}, \bibinfo{journal}{Physical Review B}
  \textbf{\bibinfo{volume}{103}}, \bibinfo{pages}{245122}
  (\bibinfo{year}{2021}).

\bibitem[{\citenamefont{Liu et~al.}(2023)\citenamefont{Liu, Huo, Huang, Huang,
  Sun, Chen, Xu, Yin, Li, and Wang}}]{Liu2023}
\bibinfo{author}{\bibfnamefont{H.}~\bibnamefont{Liu}},
  \bibinfo{author}{\bibfnamefont{M.}~\bibnamefont{Huo}},
  \bibinfo{author}{\bibfnamefont{C.}~\bibnamefont{Huang}},
  \bibinfo{author}{\bibfnamefont{X.}~\bibnamefont{Huang}},
  \bibinfo{author}{\bibfnamefont{H.}~\bibnamefont{Sun}},
  \bibinfo{author}{\bibfnamefont{L.}~\bibnamefont{Chen}},
  \bibinfo{author}{\bibfnamefont{J.}~\bibnamefont{Xu}},
  \bibinfo{author}{\bibfnamefont{W.}~\bibnamefont{Yin}},
  \bibinfo{author}{\bibfnamefont{R.}~\bibnamefont{Li}}, \bibnamefont{and}
  \bibinfo{author}{\bibfnamefont{M.}~\bibnamefont{Wang}},
  \bibinfo{journal}{arXiv:2303.00597}  (\bibinfo{year}{2023}).

\bibitem[{\citenamefont{Huang et~al.}(2023)\citenamefont{Huang, Cheng, Zhang,
  Jiang, Li, Huo, Liu, Huang, Liang, Chen et~al.}}]{huang2023}
\bibinfo{author}{\bibfnamefont{C.}~\bibnamefont{Huang}},
  \bibinfo{author}{\bibfnamefont{B.}~\bibnamefont{Cheng}},
  \bibinfo{author}{\bibfnamefont{Y.}~\bibnamefont{Zhang}},
  \bibinfo{author}{\bibfnamefont{L.}~\bibnamefont{Jiang}},
  \bibinfo{author}{\bibfnamefont{L.}~\bibnamefont{Li}},
  \bibinfo{author}{\bibfnamefont{M.}~\bibnamefont{Huo}},
  \bibinfo{author}{\bibfnamefont{H.}~\bibnamefont{Liu}},
  \bibinfo{author}{\bibfnamefont{X.}~\bibnamefont{Huang}},
  \bibinfo{author}{\bibfnamefont{F.}~\bibnamefont{Liang}},
  \bibinfo{author}{\bibfnamefont{L.}~\bibnamefont{Chen}}, \bibnamefont{et~al.},
  \bibinfo{journal}{Chinese Physics B} \textbf{\bibinfo{volume}{32}},
  \bibinfo{pages}{037802} (\bibinfo{year}{2023}).

\bibitem[{\citenamefont{Olmos et~al.}(2023)\citenamefont{Olmos, Chang, Mishra,
  Zope, Baruah, Liu, Petrovic, and Singamaneni}}]{olmos2023}
\bibinfo{author}{\bibfnamefont{R.}~\bibnamefont{Olmos}},
  \bibinfo{author}{\bibfnamefont{P.-H.} \bibnamefont{Chang}},
  \bibinfo{author}{\bibfnamefont{P.}~\bibnamefont{Mishra}},
  \bibinfo{author}{\bibfnamefont{R.~R.} \bibnamefont{Zope}},
  \bibinfo{author}{\bibfnamefont{T.}~\bibnamefont{Baruah}},
  \bibinfo{author}{\bibfnamefont{Y.}~\bibnamefont{Liu}},
  \bibinfo{author}{\bibfnamefont{C.}~\bibnamefont{Petrovic}}, \bibnamefont{and}
  \bibinfo{author}{\bibfnamefont{S.~R.} \bibnamefont{Singamaneni}},
  \bibinfo{journal}{The Journal of Physical Chemistry C}
  \textbf{\bibinfo{volume}{127}}, \bibinfo{pages}{10324}
  (\bibinfo{year}{2023}).

\bibitem[{\citenamefont{Wang et~al.}(2022)\citenamefont{Wang, Wang, He, Zhou,
  An, Zhang, Zhou, Han, Chen, Zhou et~al.}}]{wang2022}
\bibinfo{author}{\bibfnamefont{J.}~\bibnamefont{Wang}},
  \bibinfo{author}{\bibfnamefont{S.}~\bibnamefont{Wang}},
  \bibinfo{author}{\bibfnamefont{X.}~\bibnamefont{He}},
  \bibinfo{author}{\bibfnamefont{Y.}~\bibnamefont{Zhou}},
  \bibinfo{author}{\bibfnamefont{C.}~\bibnamefont{An}},
  \bibinfo{author}{\bibfnamefont{M.}~\bibnamefont{Zhang}},
  \bibinfo{author}{\bibfnamefont{Y.}~\bibnamefont{Zhou}},
  \bibinfo{author}{\bibfnamefont{Y.}~\bibnamefont{Han}},
  \bibinfo{author}{\bibfnamefont{X.}~\bibnamefont{Chen}},
  \bibinfo{author}{\bibfnamefont{J.}~\bibnamefont{Zhou}}, \bibnamefont{et~al.},
  \bibinfo{journal}{Physical Review B} \textbf{\bibinfo{volume}{106}},
  \bibinfo{pages}{045106} (\bibinfo{year}{2022}).

\bibitem[{\citenamefont{Pan et~al.}(2015)\citenamefont{Pan, Chen, Liu, Feng,
  Wei, Zhou, Chi, Pi, Yen, Song et~al.}}]{pan2015}
\bibinfo{author}{\bibfnamefont{X.-C.} \bibnamefont{Pan}},
  \bibinfo{author}{\bibfnamefont{X.}~\bibnamefont{Chen}},
  \bibinfo{author}{\bibfnamefont{H.}~\bibnamefont{Liu}},
  \bibinfo{author}{\bibfnamefont{Y.}~\bibnamefont{Feng}},
  \bibinfo{author}{\bibfnamefont{Z.}~\bibnamefont{Wei}},
  \bibinfo{author}{\bibfnamefont{Y.}~\bibnamefont{Zhou}},
  \bibinfo{author}{\bibfnamefont{Z.}~\bibnamefont{Chi}},
  \bibinfo{author}{\bibfnamefont{L.}~\bibnamefont{Pi}},
  \bibinfo{author}{\bibfnamefont{F.}~\bibnamefont{Yen}},
  \bibinfo{author}{\bibfnamefont{F.}~\bibnamefont{Song}}, \bibnamefont{et~al.},
  \bibinfo{journal}{Nature communications} \textbf{\bibinfo{volume}{6}},
  \bibinfo{pages}{7805} (\bibinfo{year}{2015}).

\bibitem[{\citenamefont{Kang et~al.}(2015)\citenamefont{Kang, Zhou, Yi, Yang,
  Guo, Shi, Zhang, Wang, Zhang, Jiang et~al.}}]{kang2015}
\bibinfo{author}{\bibfnamefont{D.}~\bibnamefont{Kang}},
  \bibinfo{author}{\bibfnamefont{Y.}~\bibnamefont{Zhou}},
  \bibinfo{author}{\bibfnamefont{W.}~\bibnamefont{Yi}},
  \bibinfo{author}{\bibfnamefont{C.}~\bibnamefont{Yang}},
  \bibinfo{author}{\bibfnamefont{J.}~\bibnamefont{Guo}},
  \bibinfo{author}{\bibfnamefont{Y.}~\bibnamefont{Shi}},
  \bibinfo{author}{\bibfnamefont{S.}~\bibnamefont{Zhang}},
  \bibinfo{author}{\bibfnamefont{Z.}~\bibnamefont{Wang}},
  \bibinfo{author}{\bibfnamefont{C.}~\bibnamefont{Zhang}},
  \bibinfo{author}{\bibfnamefont{S.}~\bibnamefont{Jiang}},
  \bibnamefont{et~al.}, \bibinfo{journal}{Nature Communications}
  \textbf{\bibinfo{volume}{6}}, \bibinfo{pages}{7804} (\bibinfo{year}{2015}).

\bibitem[{\citenamefont{Yang et~al.}(2022)\citenamefont{Yang, Liu, Chen, Liu,
  Zhang, Yu, Zhang, Sun, Uwatoko, Dong et~al.}}]{yang2022}
\bibinfo{author}{\bibfnamefont{P.~T.} \bibnamefont{Yang}},
  \bibinfo{author}{\bibfnamefont{Z.~Y.} \bibnamefont{Liu}},
  \bibinfo{author}{\bibfnamefont{K.~Y.} \bibnamefont{Chen}},
  \bibinfo{author}{\bibfnamefont{X.~L.} \bibnamefont{Liu}},
  \bibinfo{author}{\bibfnamefont{X.}~\bibnamefont{Zhang}},
  \bibinfo{author}{\bibfnamefont{Z.~H.} \bibnamefont{Yu}},
  \bibinfo{author}{\bibfnamefont{H.}~\bibnamefont{Zhang}},
  \bibinfo{author}{\bibfnamefont{J.~P.} \bibnamefont{Sun}},
  \bibinfo{author}{\bibfnamefont{Y.}~\bibnamefont{Uwatoko}},
  \bibinfo{author}{\bibfnamefont{X.~L.} \bibnamefont{Dong}},
  \bibnamefont{et~al.}, \bibinfo{journal}{Nature Communications}
  \textbf{\bibinfo{volume}{13}}, \bibinfo{pages}{2975} (\bibinfo{year}{2022}).

\end{thebibliography}
\end{document}